\documentclass[conference]{IEEEtran}
\IEEEoverridecommandlockouts
\usepackage{cite}
\usepackage{amsmath,amssymb,amsfonts}
\usepackage{algorithmic}
\usepackage{graphicx}
\usepackage{textcomp}
\usepackage{xcolor}
\usepackage{hyperref}
\usepackage{multirow}
\usepackage{booktabs}
\usepackage{algorithm}
\usepackage{algorithmic}
\usepackage{subfigure} 
\def\BibTeX{{\rm B\kern-.05em{\sc i\kern-.025em b}\kern-.08em
    T\kern-.1667em\lower.7ex\hbox{E}\kern-.125emX}}
\begin{document}

\title{DPATD: Dual-Phase Audio Transformer for Denoising\\
}

\author{\IEEEauthorblockN{Junhui Li\IEEEauthorrefmark{1}, Pu Wang\IEEEauthorrefmark{2}, Jialu Li\IEEEauthorrefmark{3}, Xinzhe Wang\IEEEauthorrefmark{4}, Youshan Zhang \IEEEauthorrefmark{5}} 
\IEEEauthorblockA{\IEEEauthorrefmark{1}\IEEEauthorrefmark{2}Department of Mathematics, School of Science, University of Science and Technology, Liaoning, Anshan, China\\ \IEEEauthorrefmark{4}School of business administration, University of Science and Technology, Liaoning, Anshan, China\\
Email: \IEEEauthorrefmark{1}Junhui\_lee@foxmail.com, \IEEEauthorrefmark{2}120203803006@stu.ustl.edu.cn, \IEEEauthorrefmark{4}1946378145@qq.com}
\IEEEauthorblockA{\IEEEauthorrefmark{3}School of public policy, Cornell University, Ithaca, NY, USA. Email: jl4284@cornell.edu}
\IEEEauthorblockA{\IEEEauthorrefmark{5}Department of Artificial Intelligence and Computer Science,  Yeshiva University, New York, NY, USA\\
Email: youshan.zhang@yu.edu}}

\maketitle

\begin{abstract}
Recent high-performance transformer-based speech enhancement models demonstrate that time domain methods could achieve similar performance as time-frequency domain methods. However, time-domain speech enhancement systems typically receive input audio sequences consisting of a large number of time steps, making it challenging to model extremely long sequences and train models to perform adequately. In this paper, we utilize smaller audio chunks as input to achieve efficient utilization of audio information to address the above challenges. We propose a dual-phase audio transformer for denoising (DPATD), a novel model to organize transformer layers in a deep structure to learn clean audio sequences for denoising. DPATD splits the audio input into smaller chunks, where the input length can be proportional to the square root of the original sequence length. Our memory-compressed explainable attention is efficient and converges faster compared to the frequently used self-attention module. Extensive experiments demonstrate that our model outperforms state-of-the-art methods.
\end{abstract}


\section{Introduction}
\label{sec:intro}

Speech signals are inevitably accompanied by various types of background noise in daily environments, such as automatic speech recognition systems, hearing aids, vehicles and mobile phones, aircraft cockpits, and multi-party conferencing devices~\cite{jiang2022speech}. In the field of speech communication, constructing efficient models to eliminate background noise is still a difficult challenge. The goal of speech enhancement learning is to find a transformation that makes clean speech readily available from the original audio. Recent achievements in transfer learning from extensive generative language models serve as a major source of inspiration for our work. There are two main challenges: (1) audio signals are continuous while textual representations are discrete; and (2) the decoder is responsible for generating text that is very different from traditional speech representation. Our efforts focus on directly creating an audio-denoising model that can learn efficiently and applied to a variety of datasets.

The transformer was originally proposed for natural language processing (NLP) tasks~\cite{vaswani2017attention} and has recently become popular in the fields of computer vision (CV)~\cite{chen2021visformer} and audio processing (AP)~\cite{subakan2021attention}. The transformer model can effectively solve the long-term dependency problem and can run well in parallel, showing good performance on many natural language processing tasks. Transformer-based methods also show promising performance in audio denoising. 
Yu et al.~\cite{yu2022dbt} proposed a dual-branch federative magnitude and phase estimation framework, named DBT-Net, for monaural speech enhancement, aiming at recovering the coarse- and fine-grained regions of the overall spectrum in parallel. Wang et al.~\cite{wang2021tstnn} proposed a two-stage transformer neural network for end-to-end speech denoising in the time domain. Yu et al.~\cite{yu2022setransformer} proposed a cognitive computing-based speech enhancement model termed SETransformer to improve speech quality in unknown noisy environments, which takes advantage of the LSTM and multi-head attention mechanisms. Dand et al.~\cite{dang2022dpt} proposed a dual-path transformer-based full-band and sub-band fusion network (DPT-FSNet) for speech enhancement in the frequency domain.

Speech typically requires long-range sequence modeling, convolutional neural networks necessitate a greater number of convolutional layers to expand the receptive field and thus continuously increase model complexity. In many natural language processing and visual tasks, self-attention transformer-based neural networks outperform deep learning models, which are constructed based on convolutional neural networks (CNNs)~\cite{cao2023swin}. Additionally, RNN models can be commonly utilized for modeling long-term sequences with sequential information, such as the Long Short-Term Memory (LSTM) and Gated Recurrent Unit (GRU)~\cite{luo2020dual}. However, RNN-based models are unable to be processed in parallel due to the temporal structure of RNNs, leading to suboptimal computational efficiency and increased complexity. Recent studies have demonstrated that a self-attention methodology can be used for audio denoising, which will relieve the challenges associated with modeling long-sequence speech signals~\cite{subakan2021attention}. The ability to learn efficiently from raw audio is crucial for constructing speech models for speech enhancement.

Inspired by the capability of the transformer in sequence modeling, we aim to develop an efficient transformer approach that enhances audio quality by learning general and meaningful speech features. We propose a dual-phase audio transformer for denoising (DPATD) that incorporates explainable attention and memory-compressed attention. The intuition is to divide the input audio into shorter chunks and interleave dual-phase transformers, a local-chunk transformer and a global-chunk transformer for local and global modeling, respectively. Additionally, to solve low-occupancy or unneeded shared memory reads and writes on the GPU, we employ FlashAttention-2, a unique attention algorithm with superior work partitioning~\cite{dao2023flashattention}. The contributions of this work are threefold: 
\begin{enumerate}
    \item We propose a dual-phase audio transformer that facilitates the modeling of audio signals by organizing transformer layers. It first segments the input audio into shorter chunks and employs dual-phase transformers in an interleaved manner, namely, a local-chunk transformer for local modeling and a global-chunk transformer for global modeling.

    \item We develop a novel transformer block that combines explainable attention and memory-compressed attention to conserve memory resources and provide interpretable attention maps that are noise-resistant and align with informative input patterns naturally.
    
    \item Our extensive experiments on two benchmark datasets demonstrate that the DPATD outperforms state-of-the-art methods across all evaluation criteria while maintaining a relatively low model complexity.
\end{enumerate}

\begin{figure*}
  \centering
   \includegraphics[width=0.7\textwidth]{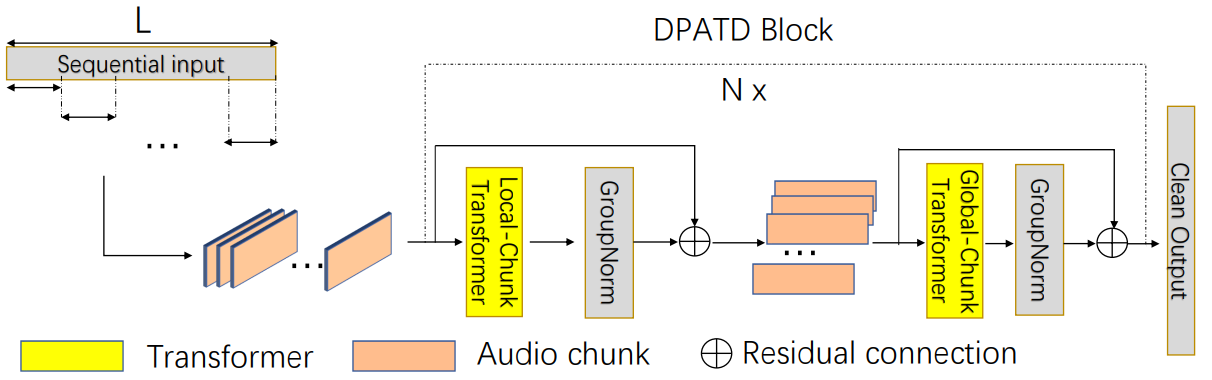}
\caption{System flowchart of the DPATD model.\textbf{ The segmentation stage splits an audio input into chunks without overlaps and concatenates them to form a 2-D tensor.} The transformer block, which consists of local-chunk and global-chunk transformers, is applied to individual chunks in parallel to process information. Multiple blocks are stacked to increase the total depth of the network. A 2-D output of the last block is converted back to an audio output.}\label{fig:over}
\end{figure*}

\section{Related work}
\textbf{Traditional speech denoising techniques.} Traditional speech denoising techniques primarily rely on statistical techniques that can be utilized to build relevant denoising models and extract clean audio from noisy input signals. The denoising performance can be improved by the Wiener filter~\cite{abd2014speech}. Linden et al.~\cite{lin2017improving} decomposed a spectral graph into a spectral basis matrix and an encoding matrix. After the different sound sources are reconstructed based on the clustering of the basis matrix and the corresponding encoding information, the noise components are removed to facilitate more accurate monitoring of biological sounds. Paliwal and Basu ~\cite{paliwal1987speech} proposed a Kalman filtering method to improve speech enhancement performance. Ali et al.~\cite{ali2017denoising} focused on the denoising of phonocardiogram (PCG) signals using different families of discrete wavelet transforms, thresholding types and techniques, and signal decomposition levels.

\textbf{Deep learning for speech enhancement.}
Different speech enhancement models based on deep neural networks have steadily taken center stage with the advancement of deep learning technology. Based on different model inputs, current voice augmentation techniques for DNNs can be broadly divided into two groups: time-domain (T) techniques and time-frequency domain (TF) techniques. Time-domain techniques employ an end-to-end model that directly estimates clean waveforms using audio data in the time domain as raw waveform inputs. The architecture foundation for time-domain approaches is WaveNet~\cite{stoller2018wave}. The majority of speech enhancement techniques currently focus on the time-frequency domain of speech. The TF techniques use the short-time Fourier transform (STFT) and the inverse short-time Fourier transform (ISTFT). The latest frequency-domain model, Band-Split RNN, explicitly splits the spectrogram of the mixture into subbands and performs interleaved band-level and sequence-level modeling for speech enhancement~\cite{yu2023efficient}.

\section{Motivation}
Most existing deep learning-based audio denoising methods study the magnitude spectrum of images for audio denoising. However, these methods can be constrained by computing power or limited filtering image regions, resulting in low denoising performance. The transformer applications in audio enhancement are still limited. Inspired by neural network approaches to text, our model encodes audio information and is trained to understand what clean audio should look like. Basically, we regard the audio signal as an "audio sequence" and further segment it into smaller chunks. The attention of each audio chunk will be calculated based on other chunks in the given audio sequence.

\section{Model}
In this section, we first review the audio denoising task, provide the motivation for our model, then conduct an in-depth analysis of the architecture of our dual-phase audio transformer for denoising (DPATD). Our model first splits the input audio into several audio chunks and encodes the audio sequence. Sequentially, generated sequence vectors are fed into a dual-phase transformer to train to minimize the difference between denoised audio and clean audio. Finally, we get the denoised audio as shown in Fig.~\ref{fig:over}.

We assume that the mixture speech signal $y(t)$ is a linear sum of the clean speech signal $x(t)$ and noise $\varepsilon(t)$, and the noisy speech $y(t)$ can be typically expressed as Eq.~\eqref{eq:exp}:
\begin{equation}\label{eq:exp}
  y(t)=x(t)+\varepsilon(t).  
\end{equation}

\subsection{Segmentation}
Our DPATD framework first splits the input audio into several local chunks,then calculates both representations and their relationship. The segmentation stage encodes the audio sequence using patch embedding with the maximum audio sequence size. The smaller utterances are zero-padded to match the size of the maximum audio sequence. Given a sequence of acoustic input vectors $Y (y_1, y_2,\cdots,y_L)\in R^{1\times L}$ where $L$ is the audio sequence length, the segmentation stage splits $Y$ into chunks of length $K$ and hop size $P$. Every sample in $Y$ appears and only appears in chunks, generating $M$ equal size chunks $D_m\in R^{1\times K}, m=1,\cdots, M$. All chunks are then concatenated together to form a 2-D tensor $T=[D_1,\cdots, D_M]\in R^{K\times M}$. Positional embeddings are added to patch embeddings to preserve positional information. We use standard learnable 1D position embeddings since the audio input is an ordered sequence. The segmentation output tensor $T$ is then passed to the stack of $N$ transformer blocks. Each block converts a 2-D tensor input to another tensor with the same shape. We denote the input tensor for block $\textbf{Z}=1,\cdots,Z$ as $T_z\in R^{K \times M}$, where $T_1=T.$ The segmentation phase of the model is shown in Fig.~\ref{fig:over}.

			

\subsection{Dual-Phase Audio Transformer}
\begin{figure}

  \centering
   \includegraphics[width=0.5\textwidth]{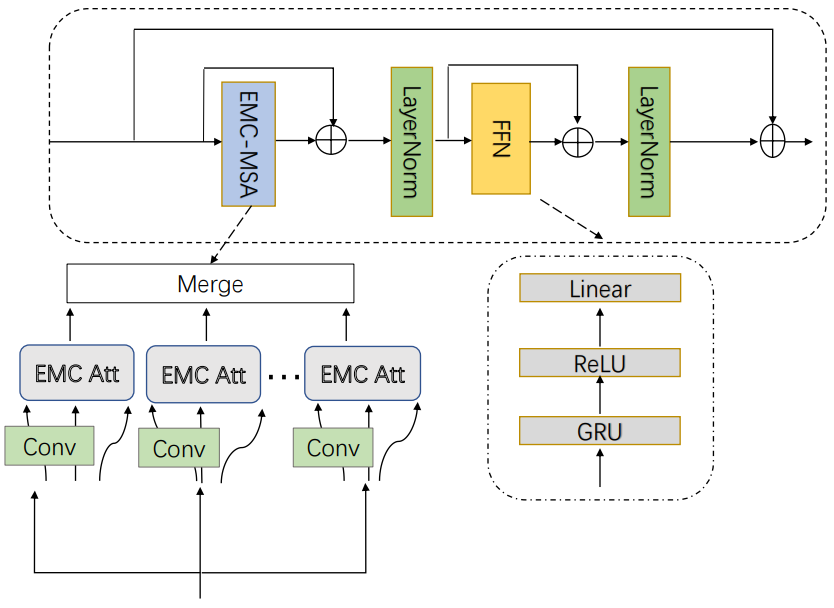}
  \vspace{-0.2cm}
\caption{The architecture of the DPAT block. \textbf{The Explainable Multi-Head Attention (E-MHA) module is capable of providing interpretable attention maps that are noise-resistant and align with informative input patterns naturally. We utilize a strided convolution to limit the dot products between $Q$ and $K$. We replace the first fully connected layer of the feed-forward network with a GRU layer to learn more positional information.}}\label{fig:block}
\end{figure}

\textbf{Basic Architecture.} Since the input noisy audio and the output enhanced audio have the same length, we introduce a simple but effective modification to the transformer for audio sequences by removing the encoder module (almost reducing model parameters by half for a given hyper-parameter set). Each transformer block contains two sub-modules. As shown in Fig.~\ref{fig:block}, the first module is a Memory-Compressed Explainable Multi-Heads Attention (MCE-MSA). This attention consists of explainable multi-heads attention and memory-compressed attention. The second module is a simple, position-wise, fully connected feed-forward network. In addition, inspired by the effectiveness of RNNs in tracking ordered sequential information, we replace the first fully connected layer of the feed-forward network with a GRU layer to learn more positional information. A residual connection was employed around each of the two sub-modules, followed by layer normalization. 

\subsection{Explainable Memory-Compressed Attention.} 

\textbf{Memory-Compressed Attention}
To handle longer sequences, we modify the multi-head attention to reduce memory usage by limiting the dot products between $Q$ and $K$ in Eq.~\eqref{eq:att}. To achieve this goal, we take advantages of a strided convolution with convolution kernels of size 3 with stride 3. Since the memory cost of attention is constant for each block, this alteration allows us to maintain the linear relationship between the number of activations and the length of the sequence~\cite{liu2018generating}. 

\textbf{Explainable Multi-Heads Attention.} In our work, attention blocks employ $h=8$ heads (the number of parallel attention layers) and map the input $h$ times to get $Q$, $K$, and $V$ representations, respectively, as described in Eq.~\eqref{eq:qkv}. Given an input $Y$, each head $H_h$ holds an explainable attention weight $A_h\in \mathcal{R}^{N\times d}$ that represents the relative importance of input features. $A_h$ aims to learn explainable features for the output through the MCE-MSA mechanism.
\begin{equation}\label{eq:qkv}
Q_i=YW^Q_i,\ K_i=YW^K_i,\ V_i=YW^V_i,
\end{equation}
where $Y\in R^{d\times k}$ is the input with sequence of length $L$ and dimension $d$, $i=1,2,\cdots,h$ and $Q_i, K_i, V_i\in R^{l\times d/h}$ are the mapped queries, keys and values respectively. $W^Q_i, W^K_i, W^V_i\in R^{d\times d/h}$ denote the $i$-th linear transformation matrix for queries, keys, and values, respectively.

The self-attention operation is constructed by Eq.~\eqref{eq:att}. $W$ implies how much attention is paid to each token.
\begin{equation}\label{eq:att}
Att(Q,K,V)=softmax(\frac{QK^T}{\sqrt{d_k}})V=softmax(W)V.
\end{equation}

The attention weight $A$ is defined as: 
\begin{equation}\label{eq:A}
  A=\mathcal{L}(W+b)^T,  
\end{equation}
where $b$ is a trainable bias term, which is introduced as an initial alignment for the input patterns. $\mathcal{L}$ is a non-linear function that scales the L2 norm of its input. 

In follows, the self-attention feature $P$ is formally expressed as:
\begin{equation}
    P=A^{T}V,\label{eq:P}
\end{equation}

According to Eq.~\eqref{eq:A}, $\Vert A \Vert \leq 1$. There $P$ in Eq.~\eqref{eq:P} is upper-bounded as follows:
\begin{equation}
    P=\Vert A\Vert \Vert|V|\Vert cos(A,V)\leq \Vert V\Vert.\label{eq:AV}
\end{equation}
When Eq.~\eqref{eq:AV} is optimized, the attention weight $A$ is proportional to $V$. In order to achieve maximal output, $A$ is driven to align with the discriminative features in $V$, instead of the uninformative noise. Therefore, $P$ can only achieve this upper bound if all possible solutions of $ v \in V$ are encoded as eigenvectors in the weight $A$. This maximization suggests that with the attention weight $A$, we will obtain an inherently explainable decomposition of input patterns.

In our work, whole sequential explainable and memory-compressed transformer blocks are computed as:

\begin{flalign}
&S^l=MCE-MSA(Z^{l-1}),\\
&Z^l=LayerNorm(Z+S^l)\label{eq:mid},\\
&FFN(Z^l)=ReLU(GRU(Z^l)W_1+b_1)W_2+b_2,\\
&Output=LN(Z^l+FFN(Z^l)),
\end{flalign}
where $LayerNorm(\cdot)$ is the LayerNorm layer, $FFN(\cdot)$ denotes the output of the position-wise feed-forward network, and $W_1\in R^{d_{ff}\times d}, b_1\in R^d$ and $d_ff=4\times d$. 

\subsection{Dual-phase Audio Transformer module}
The dual-phase audio transformer module consists of four stacked dual-chunk transformer blocks. Each block converts an input 2-D tensor into another tensor with the same shape. We propose a dual-phase transformer block based on explainable and memory-compressed attention. As shown in Fig.\ref{fig:over}, it has a local-chunk transformer and a global-chunk transformer, which extract local and global information, respectively. More specifically, the input is a 2-D tensor ($[K,S]$), and the local-chunk transformer is first applied to individual chunks to parallelly process inter information, which performs on the last dimension $F$ of the input tensor. Then, the global-chunk transformer is used to fuse the information of the output from the local-chunk transformer to learn global dependency, which is implemented on the dimension of the tensor. Besides, each transformer is followed by the group normalization operation and utilizes residual connections.

\textbf{Decoder.} We use the patch-expanding layer in the decoder to upsample the extracted deep features. The 2-D convolution with a filter size of (1, 1) recovers the channel dimension of the enhanced speech feature into 1 and produces the enhanced speech waveform by an overlap-add method.

\textbf{Loss Function.} The time-domain loss is based on the mean square error (MSE) between the input clean audio $(x_1,x_2,\cdots,x_N)$ and the predicted audio$(\hat{x}_1,\hat{x}_2,\cdots, \hat{x}_N)$. The model is optimized by minimizing the MSE, which is defined as:
\begin{equation}\label{eq:loss}
    Loss=\frac{1}{N}\sum_{i-1}^{N-1}(x_i-\hat{x}_i)^2,
\end{equation}
where $N$ denotes the number of samples.

The overall training algorithm is shown in Alg.~\ref{alg:DPATD}.
\begin{algorithm}[ht]
   \caption{DPATD: Dual-Phase Audio Transformer for Denoising. Batch of audio input: $B(Y)=\{Y^1,...,Y^{n_B}  \}$, and their clean audio input $B(X)=\{X^1,...,X^{n_B}\}$, where ${n_B}$ is the total number of batch. $I$ is the number of iterations }
   \label{alg:DPATD}
\begin{algorithmic}[1]
   \STATE {\bfseries Input:} Mixture audio signals $Y=\{y_i\}_{i=1}^{N}$ and clean audio input $X=\{x_i\}_{i=1}^{N}$, where $N$ is the total sample number of audios.
   \STATE {\bfseries Output:} Denoised audio signals: $\hat{X}$
   \STATE Segmentation splits the input audio into a 2-D tensor $T=[D_1,\cdots, D_M]$ using 1-D CNNs.
   \FOR{$iter =1$ {\bfseries to} $I$}
   \STATE Derive batch-wise data: $Y^k$ and $ X^k$ sampled from $B(Y)$ and $B(X)$
   \STATE Optimize our audio generation model DPATD using Eq.~\eqref{eq:loss}
   \ENDFOR
   \STATE Output the denoised audio signals 
\end{algorithmic}
\end{algorithm}

 \section{Experimental Procedures}


\subsection{Dataset}
\textbf{VCTK+DEMAND dataset.} We validate the effectiveness of our proposed model on a standard speech dataset ~\cite{valentini2016speech}. The clean speech datasets are selected from the Voice Bank Corpus, including a training set of 11572 utterances from 28 speakers and a test set of 872 utterances from 2 speakers. 

\textbf{BirdSoundsDenoising.} This dataset uses a variety of natural noises, such as wind, rain, waterfalls, etc., in place of the normal intentionally generated noise~\cite{zhang2023birdsoundsdenoising}. In particular, the dataset contains 14,120 audios from one second to fifteen seconds and is a large-scale dataset of bird sounds collected, containing 10,000/1,400/2,720 in training, validation, and testing sets, respectively. 

\begin{figure}[t]
	\centering
	\includegraphics[width=0.5\textwidth]{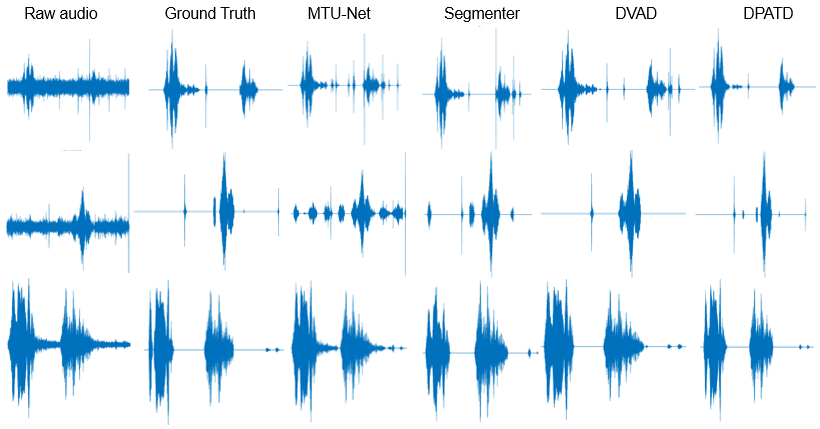}
	\caption{Denoising results comparisons. Raw audio is the original noisy audio. }\label{Fig:audiocom}
 	\vspace{-0.3cm}
\end{figure}

\subsection{Implementation details} We train the DPATD model for 100 epochs on mini-batches of 8 random samples, where the segment chunk size is set to 1000. We used the Adam optimization scheme with a maximum learning rate of 2.5e-4. Since the layernorm layer is extensively  used throughout the model, a simple weight initialization of $N(0, 0.02)$ was adequate. For the activation function, we used the Gaussian Error Linear Unit (GELU). We used the learned position embeddings instead of the sinusoidal version proposed in the original work. With a single NVIDIA GTX 3060 GPU and PyTorch to implement our model, it took around 180 hours of GPU time to train our model. The smaller utterances in a batch are zero-padded to the largest size of utterances. A dynamic strategy is used to adjust the learning rate during the training stage~\cite{chen2020dual}. 

\textbf{Model configurations.} The configuration of our model structure partially follows the original transformer architecture. We trained a 12-layer decoder-only transformer with self-attention heads (1000 dimensional states and 12 attention heads). Here, we focus on explainable and memory compression attention to effectively compute and conserve memory. For the position-wise feed-forward networks, we used 4000-dimensional inner states. Consider the sum of the input audio length in a signal block denoted by $K\times S$ where the hop size is set to be the same as the embedding size. It is simple to see that $S=[L/K]$ where $[\cdot]$ is the ceiling function. To achieve the minimum total input length $K+S=K+[L/K]$, $K$ is selected such that $K\approx sqrt(5L)$. This gives us sublinear input length$(O(sqrt(L))$ rather than the original linear input length $(O(L))$.



\begin{table}[t]

\begin{center}
\caption{Results comparisons of different methods ($F1, IoU$, and $Dice$ scores are multiplied by 100. ``$-$" means not applicable. }
\label{tabcom}
\vspace{-0.2cm}
\setlength{\tabcolsep}{+0.8mm}{
\begin{tabular}{lllll|lllllllll}
\hline \label{tab:md}
 \multirow{4}{*}{Networks} & \multicolumn{4}{c}{Validation} & \multicolumn{4}{c}{Test} \\
 \cmidrule{2-9}
& $F1$ & $IoU$  & $Dice$ & $SDR$ & $F1$ & $IoU$  & $Dice$ & $SDR$ \\
\hline
U$^2$-Net~\cite{qin2020u2}  &60.8 &45.2 &60.6 &7.85 & 60.2  &44.8 &59.9 & 7.70\\
MTU-NeT~\cite{wang2022mixed}  &69.1 &56.5 &69.0  &8.17 & 68.3  &55.7 & 68.3 &7.96  \\
Segmenter~\cite{strudel2021segmenter} & 72.6  & 59.6 & 72.5 & 9.24 & 70.8 & 57.7 & 70.7 & 8.52   \\
SegNet~\cite{badrinarayanan2017segnet}  &77.5 &66.9 &77.5 & 9.55&76.1 &65.3 &76.2 & 9.43 \\
DVAD~\cite{zhang2023birdsoundsdenoising} & 82.6  & 73.5 & 82.6 & 10.33  & 81.6 & 72.3 & 81.6 & 9.96 \\
R-CED~\cite{park2016fully} & $-$ & $-$ & $-$ &2.38     &$-$ &$-$&$-$ & 1.93  \\
Noise2Noise~\cite{kashyap2021speech}  & $-$ & $-$ & $-$ & 2.40&$-$ &$-$&$-$ &1.96\\
TS-U-Net~\cite{moliner2022two}  & $-$ & $-$ & $-$ & 2.48&$-$ &$-$&$-$ &1.98\\
\textbf{DPATD} & $-$  & $-$ & $-$ &  \textbf{10.49}  & $-$ & $-$ & $-$ & \textbf{10.43} \\
\hline
\end{tabular}}
\end{center}
\vspace{-.6cm}
\end{table}

\subsection{Result}
Tab.~\ref{tab:voice} shows the comparison results of the VCTK+DEMAND dataset. Our model surpasses most waveform-based approaches now in use in terms of the PESQ score and achieves performance that is equivalent to other methods in other evaluation metrics. For the BirdSoundsDenoising dataset, we report the performance of eight state-of-the-art baselines. The results are shown in Tab.~\ref{tabcom}, where the bold text indicates the best outcomes for each statistic. The results demonstrate that our model outperforms other state-of-the-art methods in terms of the SDR. Results of F1, IoU, and Dice are not included because these metrics are used for the audio image segmentation task~\cite{li2023deeplabv3+,zhang23p_interspeech}. The comparisons of raw bird audio, ground truth labeled denoised audio, and denoised audio of other models are shown in Fig.~\ref{Fig:audiocom}. Additionally, our model bears more resemblance to the labeled denoised signal. As a consequence, our model enhances the audio-denoising capabilities of the BirdSoundDenoising dataset.

\begin{table}
  \caption{Comparison results on the VoiceBank-DEMAND dataset. ``$-$" means not applicable.}
  \label{tab:voice}
  \centering
\setlength{\tabcolsep}{+1mm}{
  \begin{tabular}{lclllll}
    \toprule
    Methods  & Domain & PESQ & STOI & CSIG & CBAK & COVL  \\
    \hline

PGGAN~\cite{li2022perception} &T &2.81 &0.944& 3.99& 3.59 &3.36 \\
DCCRGAN~\cite{huang2022dccrgan} &TF &2.82 &0.949 &4.01& 3.48 &3.40 \\
S-DCCRN~\cite{lv2022s} &TF &2.84 &0.940 &4.03 &2.97 &3.43 \\
TSTNN~\cite{wang2021tstnn} & T & 2.96 & 0.950 & 4.33 & 3.53 & 3.67\\
PHASEN~\cite{yin2020phasen} &TF &2.99 &$-$ &4.18& 3.45& 3.50 \\
DEMUCS~\cite{defossez2020real} & T & 3.07 &0.95& 4.31 & 3.40 & 3.63 \\
SE-Conformer~\cite{kim2021se} & T & 3.13 &0.95& 4.45 & 3.55 & 3.82 \\
MetricGAN+~\cite{fu2021metricgan+} &TF &3.15 &0.927& 4.14& 3.12 &3.52\\
MANNER~\cite{park2022manner} & T & 3.21 & 0.950 & 4.53 & 3.65 & 3.91\\
CMGAN~\cite{cao2022cmgan} & T & 3.41 &0.96& 4.63 & 3.94 & 4.12 \\
\hline
DPATD & T & \textbf{3.55} & \textbf{0.97} & \textbf{4.78} & \textbf{3.96} & \textbf{4.22} \\
\hline
  \end{tabular}}
  \vspace{-0.3cm}
\end{table}
\textbf{Ablation study.}
First, we examine the performance of our method by comparing it with different transformer layers and the effect of chunks in segmentation. In order to further demonstrate the effectiveness of our proposed transformer block, we also designed another architecture for comparison. In this architecture, we use different transformer blocks rather than the 12 blocks in the DPATD, and we increase or reduce the number of heads. In addition, we also set chunk sizes at 500 and 2000, while only 1000 in our DPATD. From Tab.\ref{tab:layers}, a 12-transformer-block DPATD has better scores than other models. As shown in Tab.\ref{tab:chunks}, the chunk has only a slight influence on the performance, and we use chunk=1000 by default for its efficiency.

\begin{table}
  \setlength{\tabcolsep}{1mm}
  \centering
  \caption{Effect of layers and chunks in transformer block in the DPATD.}
  \vspace{-0.3cm}
  \subtable[]{
    \begin{tabular}{ccccc}
        \toprule
         Heads & 6 & 8 & 12 & 16 \\
        \hline
        PESQ&2.99&3.23&3.45&3.15\\
        \hline
        \label{tab:layers}
    \end{tabular}}
  \subtable[]{
    \begin{tabular}{ccccc}
       \toprule
        Chunks & 500 & 1000 & 2000  \\
        \hline
        PESQ&3.26&3.45&3.38\\
        \hline
        \label{tab:chunks}
    \end{tabular}
     }
\vspace{-0.5cm}
\end{table}


\section{Conclusion}
In this study, we present a framework for using a dual-phase audio transformer for denoising (DPATD) to provide robust speech enhancement. The DPATD splits the audio input into non-overlapping chunks in the segmentation stage, which are then passed as input to the transformer model. In a DPAT block, the local-chunk transformer and global-chunk transformer process the local chunks and all the chunks, respectively. We modified the transformer model using explainable multi-head attention and memory-compressed attention. Extensive experiments on datasets have demonstrated the effectiveness and superiority of the proposed DPATD architecture. Finally, our method is still computationally demanding, and future directions of the work could improve on these limitations.
\bibliographystyle{unsrt}
\bibliography{refs}

\end{document}